\documentclass[superscriptaddress,preprintnumbers,reprint,prl]{revtex4-1}

\usepackage[totalheight = 23cm, totalwidth = 17cm]{geometry}
\usepackage{amssymb,amsmath,amsfonts,amsbsy,graphicx}
\usepackage{bm,color}
\usepackage{hyperref}

\newcommand{\mpl}{m_{\rm Pl}}
\newcommand{\sn}{{\rm sn}}
\newcommand{\csch}{{\rm csch}}
\newcommand{\calL}{{\cal L}}
\newcommand{\calO}{{\cal O}}
\newcommand{\calR}{{\cal R}}

\begin{document}

\title{Natural Cliff Inflation}

\author{Jinn-Ouk Gong }
\affiliation{Asia Pacific Center for Theoretical Physics, Pohang 37673, Korea}
\affiliation{Department of Physics, Postech, Pohang 37673, Korea}

\author{Chang Sub Shin}
\affiliation{Asia Pacific Center for Theoretical Physics, Pohang 37673, Korea}
\affiliation{Center for Theoretical Physics of the Universe, 
Institute for Basic Science, Daejeon 34051, Korea}

\begin{abstract}

We propose a novel scenario of inflation, in which the inflaton is identified as the lightest mode of an angular field in a compactified fifth dimension. The periodic effective potential exhibits exponentially flat plateaus, so that a sub-Planckian field excursion without hilltop initial conditions is naturally realized. We can obtain consistent predictions with observations on the spectral index and the tensor-to-scalar ratio.

\end{abstract}

\preprint{APCTP Pre2017-020, CTPU-17-39}

\maketitle

\section{Introduction}


Currently, the cosmic microwave background (CMB) is observed to be homogeneous and isotropic, and its temperature fluctuations indicate a nearly scale-invariant power spectrum of the adiabatic Gaussian primordial perturbation~\cite{Ade:2015xua}. These are consistent with the predictions of inflation~\cite{Ade:2015lrj}, which is thus regarded as the leading candidate for the evolution of the very early universe before the standard hot big bang~\cite{Guth:1980zm,Linde:1981mu,Albrecht:1982wi}. Most models of inflation introduce a scalar field, the inflaton, to drive inflation in different extensions of the standard model of particle physics (SM)~\cite{Lyth:1998xn}. However, constructing a realistic model of inflation in particle physics is challenging. For example, inflation models in supergravity receive typically $\calO(H^2)$ corrections to the inflaton mass, with $H$ being the Hubble parameter during inflation, so that the flatness of the inflaton potential is completely spoiled~\cite{Copeland:1994vg}.


An inflaton protected by symmetry is thus a good candidate to maintain a flat potential to support long enough inflation. Natural inflation is a representative model, where the inflaton is a pseudo Nambu-Goldstone boson associated with a spontaneously broken symmetry at a scale $f$~\cite{Freese:1990rb,Adams:1992bn}. The resulting Lagrangian includes a sinusoidal periodic potential as a function of an angular variable $\theta \equiv \phi/f$ with $\phi$ being the canonically normalized field. Naively implementing natural inflation, however, requires $f \gg \mpl \equiv (8\pi G)^{-1/2} = 2.43 \times 10^{18}$ GeV so that the validity of the description of natural inflation is doubtful in the context of effective field theory. Indeed, $f \sim 5\mpl$ gives consistent predictions with the CMB observations~\cite{Ade:2015lrj}. But it seems difficult to have $f \gtrsim \mpl$ from first principles~\cite{Banks:2003sx}.


It is thus an important task, to realize natural inflation, either to obtain a super-Planckian value of $f$, at least effectively, or conversely to find working models that give consistent predictions with observations even with a sub-Planckian $f$. A number of proposals has been suggested for $f \gtrsim \mpl$, including a five-dimensional (5D) model with a gauge field~\cite{ArkaniHamed:2003wu}, aligned axion model~\cite{Kim:2004rp,Choi:2014rja} and N-flation model~\cite{Dimopoulos:2005ac}. Meanwhile, it is in general difficult to achieve a sub-Planckian $f$ and only few models are known~\cite{Germani:2010hd} since fine-tuning of initial conditions is required.


In this article, we present a new model of inflation with $f \lesssim \mpl$ by making use of a 5D angular field with appropriate boundary interactions. The hilltop of the potential becomes exponentially flat so that a sub-Planckian field excursion of the inflaton is naturally realized without severe fine tuning on the initial conditions, leading to a highly suppressed tensor-to-scalar ratio and an observationally viable value for the spectral index.

\section{Model}
\label{sec:model}

We begin with a 5D model 
\begin{equation}
ds^2 = G_{MN} dx^M dx^N = g_{\mu\nu}(x) dx^\mu dx^\nu + dy^2 \, , 
\end{equation}
where $M=\{0,1,2,3,5\}$ and $\mu=\{0,1,2,3\}$ are respectively the 5D and non-compact usual 4D coordinate indices. The fifth dimension is compactified with an orbifold symmetry $S_1/Z_2$, which has two 4D boundaries (or branes) at $y_b= 0$ and $y_b=\pi R$. We restrict $0 \lesssim y \lesssim \pi R$ so that the length of the fifth dimension is $\pi R$. This gives the relation between $\mpl$ and the 5D mass scale $M_5$ as $\mpl = \sqrt{ M_5^3 \pi R}$.

Our strategy to realize inflation is as follows. We introduce a 5D angular field $\theta(x,y)$ whose period is $2\pi$, and the corresponding 5D action $S_0[\theta]$, which generates a zero mode that we identify as a 4D massless field $\phi(x)$. Since $\phi(x)$ is massless within  $S_0[\theta]$, there is no scalar potential for $\phi(x)$. By introducing small deviations $\delta S_0[\theta]$ in the model parameters from $S_0[\theta]$ with a reasonable assumptions, $\phi(x)$ is no longer flat. This gives, after integrating out all Kaluza-Klein (KK) modes, the scalar potential of $\phi(x)$ which can drive inflation. The curvature of the potential is determined by the overlap between $S_0[\theta]$ and the zero mode localized at, say, $y_b = 0$ [see \eqref{eq:localization}].

$\theta$ has the constant shift symmetry $\theta\to \theta+ c$ as a usual Goldstone boson, and such a shift symmetry can be softly broken by a massive parameter $m$ which transforms as $m\to me^{-i c}$. With a 5D decay constant of $\theta$, $f_5$, we can write the generic 5D effective action for the 5D  pseudo Goldstone boson $\theta$ as 
\begin{equation}
S_0[\theta] =  \int d^5x f_5^3 \left[ - \frac{1}{2}(\partial_M\theta)^2 
- V_{\rm soft} \left( me^{i\theta}, m^*e^{-i\theta} \right) \right] \, .
\end{equation}
Usually a non-zero value of $m$ breaks the shift symmetry, and there is no massless mode. However if we allow boundary-localized potentials for $\theta$, then the massless mode could be generated with a non-trivial profile along the fifth dimension. Imposing a CP symmetry, the simplest example for the bulk and boundary potentials is 
\begin{eqnarray}
V_{\rm soft} &=& - \frac{1}{4} m^2\cos (2\theta) +  \alpha|m|^2  
\nonumber\\
&&- 2 m \cos\theta \Big[ \beta_{0}\delta(y) - \beta_{\pi}\delta(y-\pi R) \Big] \, ,
\end{eqnarray}
where $\alpha$, $\beta_{0}$ and $\beta_{\pi}$ are the model parameters. The phase of $m$ is completely absorbed into $\theta$. The most interesting parameter values are $\alpha=1/4$ and $\beta_0=\beta_\pi =1$, which yield a massless 4D scalar with a vanishing cosmological constant as a solution to the 5D equation of motion for the zero mode, $\partial_y\theta_0(x,y) +m \sin\theta_0(x,y)=0$. It is related to the canonically normalized scalar field $\phi(x)$ as 
\begin{equation}
\label{eq:zeromodesol}
\theta_0(x,y) = 2\tan^{-1} \left[ i e^{-my} \sn \left( \frac{\phi(x)}{2 i f} \bigg| e^{- 2m\pi R} \right) \right] \, ,
\end{equation} 
where 
\begin{equation}
f \equiv \sqrt{\frac{f_5^3}{2m} (1 -e^{-2m\pi R})} \, .
\end{equation}
Here $\sn(x|y)$ is the Jacobi elliptic function. Then, the action for the zero mode is simply the kinetic term of $\phi(x)$: $S_0[\theta_0] = - \int d^4 x \sqrt{-g} (\partial_\mu \phi)^2/2$.

\begin{figure}[h!]
 \begin{center}
  \includegraphics[width=0.34\textheight]{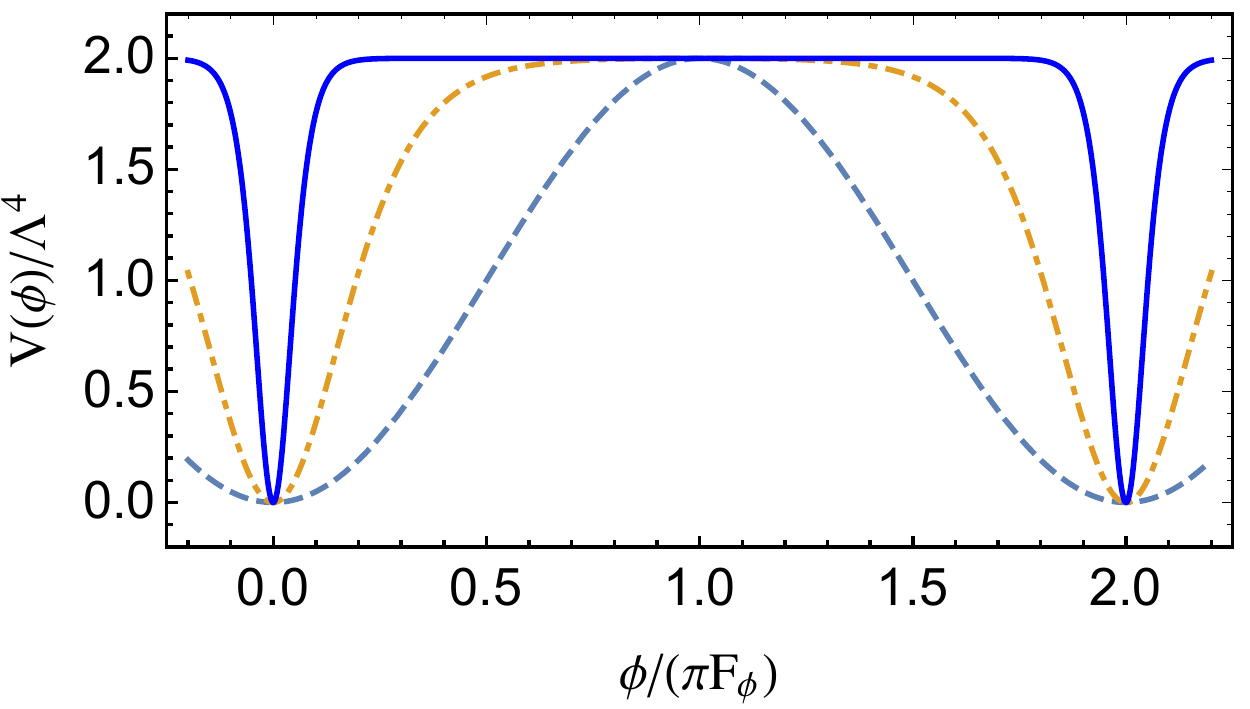}
 \end{center}
\caption{Shape of the inflaton potential $V(\phi)/\Lambda^4$ given by \eqref{eq:potential} with $mR= 0$ (dashed), $1$ (dot-dashed) and $5$ (solid). Note that the case $mR=0$ corresponds to the vanilla natural inflation with a pure sinusoidal potential.}
\label{fig:Vinf}
\end{figure}

Now we consider at $y_b=0$ a small deviation $|\delta\beta_0| = |\beta_0 - 1| \ll 1$, and correspondingly $|\delta\alpha| \ll 1$ to keep the cosmological constant vanishing. Such a deviation can be considered as introducing a small perturbation of the action, 
\begin{equation}
\label{eq:localization}
\delta S_0[\theta] = - \int d^5 x \sqrt{-g} \delta(y)\,f_5^3 m \delta\beta_0 \big[ 1- \cos\theta(x, y) \big] \, .
\end{equation}
After integrating out all KK modes, the canonical 4D field $\phi(x)$ acquires a potential:
\begin{equation}
\label{eq:potential}
V(\phi) = - \Lambda^4\frac{\sn^2\left(\dfrac{\phi}{2i f}\bigg|e^{- 2m\pi R}\right)}
{1 - \sn^2\left(\dfrac{\phi}{2i f}\bigg| e^{- 2m\pi R}\right)} \, ,
\end{equation}
where $\Lambda^4 \equiv 2 f_5 m \delta\beta_0$. We see that \eqref{eq:potential} is a periodic function of $\phi$. In Figure~\ref{fig:Vinf}, the potential is shown for different values of $m R$ in the unit of the half period, $\pi F_\phi$, given by
\begin{equation}
\pi F_\phi = 2 f K(1- e^{-2m\pi R}) \, , 
\end{equation}
where $K(x)$ is the complete elliptic integral of the first kind, which approaches $\pi/2$ ($m\pi R$) for $mR \ll 1$ ($mR \gg 1$). Note that depending on the value of $mR$ the shape of the potential is exponentially sensitive in such a way that the hilltop becomes an exponentially flat plateau, while around the minimum $V(\phi)$ exhibits a steep, cliff-like shape. For successful inflation, we take $mR=\calO(10)$, so we can naturally obtain a flat and periodic inflaton potential.

There is a clear geometric interpretation of the potential: it is the result of the overlap of the wavefunction between the exponentially localized zero mode \eqref{eq:zeromodesol}, whose position along the fifth dimension depends on the vacuum value of $\phi$, and the perturbed action \eqref{eq:localization} which breaks the shift symmetry of the zero mode. A more detailed discussion about the zero mode localization could be found in~\cite{Choi:2017ncj} in the context of continuum clockwork.

\section{Predictions}
\label{sec:predictions}

Given the potential \eqref{eq:potential}, from the slow-roll parameters
\begin{equation}
\epsilon \equiv \frac{\mpl^2}{2} \left( \frac{V'}{V} \right)^2 
\quad \text{and} \quad
\eta \equiv \mpl^2 \frac{V''}{V} \, ,
\end{equation}
where a prime denotes a derivative with respect to $\phi$, we can first check the number of $e$-folds for a given initial vacuum expectation value $\phi_i$,
\begin{equation}
\label{eq:efold}
N = \frac{1}{\mpl} \int_{\phi_e}^{\phi_i} \frac{d\phi}{\sqrt{2\epsilon}} \, .
\end{equation}
Here $\phi_e$ satisfies $\epsilon(\phi_e)=1$. Then the amplitude $A_\calR$ and the spectral index $n_\calR$ of the power spectrum of the curvature perturbation and the tensor-to-scalar ratio $r$, given respectively by
\begin{equation}
A_\calR = \frac{V}{24\pi^2\mpl^4\epsilon} \, ,
\quad
n_\calR = 1 - 6\epsilon + 2\eta \, ,
\quad
r = 16\epsilon \, ,
\end{equation}
can be straightly computed.

While it is indeed possible to make use of \eqref{eq:potential} to obtain fully analytic results in terms of the Jacobi elliptic functions, as mentioned at the end of the previous section, with $mR=\calO(10)$ the form of the potential and the period are greatly simplified to
\begin{equation}
\label{eq:simplerV}
V(\phi) \approx \Lambda^4 \tanh^2 \left( \frac{\phi}{2f} \right)
\quad \text{for} \quad
|\phi| < \pi F_\phi \approx 2m\pi Rf \, .
\end{equation}
This simple potential gives further analytic insights. It is important to note that we can easily make the period well below $\mpl$: comparing $\pi F_\phi$ and $\mpl$, 
\begin{equation}
\pi F_\phi 
\approx \left(\frac{2 m\pi R\,f_5^3}{M_5^3}\right)^{1/2} \mpl \, ,
\end{equation}
which becomes much smaller than $\mpl$ for $f_5\ll M_5$. Also note that the same form of the potential \eqref{eq:simplerV} was studied in a different context~\cite{Kallosh:2013hoa,Kallosh:2013yoa}.

From \eqref{eq:efold} and \eqref{eq:simplerV}, we find
\begin{equation}
N = \frac{1}{2} \left( \frac{f}{\mpl} \right)^2 \left[ \cosh \left( \frac{\phi_i}{f} \right) 
- \cosh \left( \frac{\phi_e}{f} \right) \right] \, .
\end{equation}
With $\phi_e = f \sinh^{-1} \big( \sqrt{2}\mpl/f \big)$, the contribution of $\phi_e$ to $N$ is always smaller than 1, e.g. 0.866 even for $f=\mpl$. Thus $N$ is even further simplified to
\begin{equation}
N \approx \left( \frac{f}{2\mpl} \right)^2 e^{\phi_i/f} \, .
\end{equation}
In this case, we can write $n_\calR$ and $r$ in terms of $N$ as
\begin{align}
\label{eq:index}
n_\calR & = 1 - \frac{4 \big[ 1+\cosh(\phi_i/f) \big] \csch^2(\phi_i/f)}{(f/\mpl)^2}
\approx 1 - \frac{2}{N} \, ,
\\
\label{eq:ratio}
r & = \frac{32\csch^2(\phi_i/f)}{(f/\mpl)^2}
\approx \frac{8}{N^2} \left( \frac{f}{\mpl} \right)^2 \, .
\end{align}
Thus we see that comparing with the $R^2$ inflation model~\cite{Starobinsky:1980te}, one of the best fit models for the most recent CMB observations~\cite{Ade:2015lrj}, while the spectral index is the same, the tensor-to-scalar ratio is even further suppressed for a sub-Planckian value of $f$. This is because the field excursion during inflation is more sub-Planckian for smaller $f/\mpl$~\cite{Lyth:1996im}.

\begin{figure}[h!]
 \begin{center}
  \includegraphics[width=0.3\textheight]{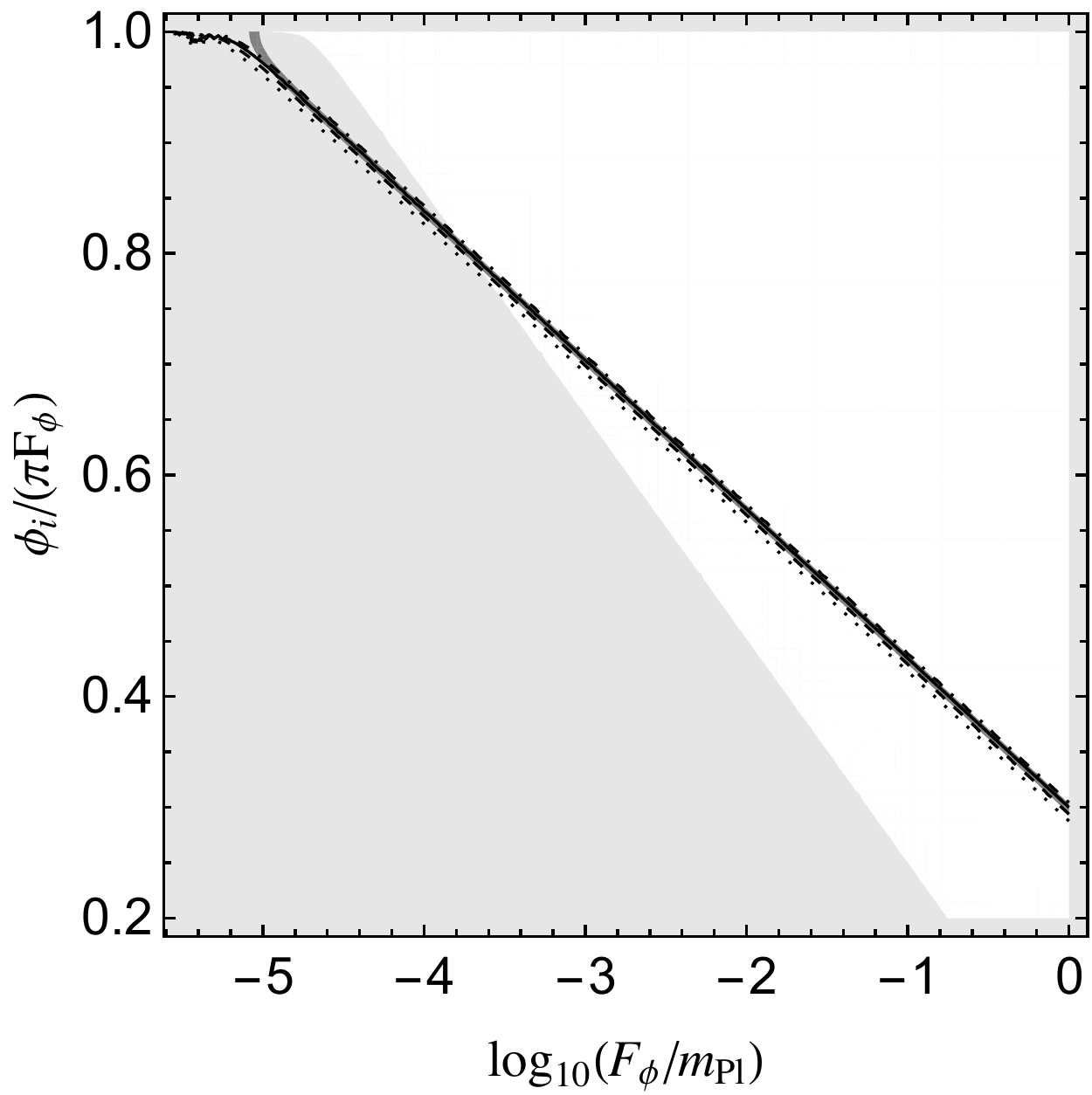}
 \end{center}
\caption{The number of $e$-folds $N$ and the specral index $n_\calR$ in the $\log_{10}(F_\phi/\mpl)$-$\phi_i/(\pi F_\phi)$ plane. We set $N=40$ (dotted), 50 (dashed), 60 (solid) and 70 (dot-dashed), and $0.96 < n_\calR < 0.97$ (dark shade). $r < 0.07$ is satisfied everywhere shown in this plot. The regime beyond the perturbative constraint $\Lambda^4/(2f_5^3m) > 0.1$ (light shade) is also shown.}
\label{fig:predictions}
\end{figure}

The observational constraints $0.96 \lesssim n_\calR \lesssim 0.97$ and $r \lesssim 0.07$ along with a few different numbers of $N$ are shown in Figure~\ref{fig:predictions} with $mR=5$. Unlike natural inflation, larger $f/\mpl$ does not necessarily lead to larger $N$. This is because as can be seen from Figure~\ref{fig:Vinf} the potential away from the hilltop is exponentially steeper than simple sinusoidal form, so that the field value on the ridge of the potential is more crucial to determine $N$. The spectral index is aligned almost parallel to $N$, since as can be read from \eqref{eq:index} in the large $e^{2\pi mR}$ limit it is entirely determined only by $N$. Since the tensor-to-scalar ratio is further suppressd by $(f/\mpl)^2$ which we take less than 1, the constraint $r < 0.07$ is satisfied in the shown parameter space.

\section{Discussions}
\label{sec:discuss}

In the standard scenario of natural inflation where inflation is driven by a simple cosine potential, $A_\calR$ is determined by two independent parameters of the potential, the axion decay constant $f$ and the energy scale $\Lambda$. To satisfy the constraint $A_\calR \approx 2.21 \times 10^{-9}$ with the preferred value $f \sim 5\mpl$, $\Lambda$ is fixed by $\Lambda \sim 10^{15-16}$ GeV. In our case, the energy scale $\Lambda$ is not an independent parameter since it should satisfy the perturbative constraint $\Lambda^4 \ll 2f_5^3m$. For $V \approx \Lambda^4$ and $\epsilon \approx (f/\mpl)^2/(2N^2)$, we can find
\begin{equation}
\label{eq:Lambda-const}
\delta\beta_0 \approx 0.0073 \left( \frac{60}{N} \right)^2 \left( \frac{10^{-3}\mpl}{\pi F_\phi} \right)^2
\left( \frac{f_5/m}{2} \right)^3 \left( \frac{mR}{5} \right)^2 \, .
\end{equation}
Note that there is no explicit hierarchy between $f_5$ and $m$, and in the above we have assumed they are not too different with a fiducial value $f_5/m = 2$. For given values of $f_5/m$ and $mR$, we can see from \eqref{eq:Lambda-const} that the perturbative constraint is always satisfied for a wide region in the parameter space. In Figure~\ref{fig:predictions}, we also present the constraint $\Lambda^4/(2f_5^3m) < 0.1$.

After inflation ends, the inflaton oscillates around the potential minimum where the effective mass is $M_\phi^2 = {\cal O}(\delta \beta m^2)$, and then decays into the SM particles and reheats the universe. The coupling to SM depends on the position of the boundary where the SM sector is localized. If SM is localized at $y_b=0$, reheating happens by the anomalous coupling like $\int d^5 x\, \sqrt{-g}\delta(y) c_\gamma \theta(x,y) F_{\mu\nu}(x)\widetilde F^{\mu\nu}(x)$. This yields the interaction Lagrangian density between the inflaton fluctuation and the gauge fields as
\begin{equation}
\delta\calL = \frac{ c_\gamma }{f} \delta\phi F\widetilde F \, .
\end{equation} 
The decay rate of the inflaton is then estimated as
\begin{equation} 
\Gamma_\phi \sim  c_\gamma^2 \frac{M_\phi^3}{f^2} 
\sim c_\gamma^2 \delta\beta^{3/2} \frac{m^4}{f_5^3} 
\lesssim \calO \left( c_\gamma^2  \delta\beta^{3/2} f_5 \right) \, .
\end{equation}
For $F_\phi \sim 10^{-3} \mpl$ and $mR = 5$, we find $f_5 \sim 10^{-3} M_5 \sim 10^{-5} \mpl$, and the reheating temperature $T_{\rm reh} = \sqrt{\Gamma_\phi\mpl} \sim 10^{13} \, {\rm GeV} (c_\gamma/0.1) (\delta \beta/0.1)$ could be large enough but safely smaller than the GUT scale. On the other hand, if SM is localized at $y_b=\pi R$, the coupling between the inflation and the gauge fields has an additional suppression factor $e^{-m\pi R}\sim 10^{-7}$ for $m R= 5$. Then $T_{\rm reh}$ is suppressed as $10^6\,{\rm GeV}(c_\gamma/0.1) (\delta \beta/0.1)$, which also could be a reasonable value. Further interesting phenomenology with such a low $T_{\rm reh}$, e.g. large isocurvature perturbation at small scales~\cite{Choi:2015yma}, may follow depending on the detail of the matter sector.

\section{Summary}
\label{sec:summary}

In this article, we have constructed a model of inflation where the zero mode of a 5D angular field $\theta$ plays the role of the inflaton. The resulting inflaton potential is exponentially flat depending on the mass parameter $m$ that softly breaks the shift symmetry of $\theta$ and the size of the compactified fifth dimension $R$. The effective 4D decay constant $f$ can be made substantially sub-Planckian, satisfying in the perturbative regime the observational constraints on the spectral index $n_\calR$ and predicting the tensor-to-scalar ratio $r$ highly suppressed.

\section{acknowledgements}

CSS thanks Kiwoon Choi for useful discussions. 
We acknowledge the support from the Korea Ministry of Education, Science and Technology, Gyeongsangbuk-Do and Pohang City for Independent Junior Research Groups at the Asia Pacific Center for Theoretical Physics. 
JG is also supported in part by a TJ Park Science Fellowship of POSCO TJ Park Foundation and the Basic Science Research Program of the National Research Foundation of Korea Research Grant 2016R1D1A1B03930408. 
CSS is supported by IBS under the project code IBS-R018-D1. 
CSS was also supported in part by the Basic Science Research Program of the National Research Foundation of Korea Starting Grant 2017R1D1A1B04032316.

\end{document}